\documentclass[12pt]{article}
\usepackage{amssymb,amsmath,amsthm,amscd,latexsym}
\usepackage{mathrsfs}
\usepackage{mathrsfs}
\usepackage{amsfonts}
\usepackage{amsmath}
\usepackage{amssymb}
\usepackage{amscd}

\renewcommand{\paragraph}{\roman{paragraph}}
\input cyracc.def

\newcommand{\F}{\mathbb{F}}

\newtheorem{theorem}{Theorem}

\newtheorem{example}{Example}

\newtheorem{lemma}{Lemma}
\newtheorem{proposition}{Proposition}

\theoremstyle{definition}

\newtheorem{remark}{Remark}

\begin{document}
\title{\bf New Classes of $p$-ary Few Weights Codes
\thanks{This research is supported by National Natural Science Foundation of China (61672036),
Technology Foundation for Selected Overseas Chinese Scholar, Ministry of Personnel of China (05015133) and
the Open Research Fund of National Mobile Communications Research Laboratory, Southeast University (2015D11) and
Key projects of support program for outstanding young talents in Colleges and Universities (gxyqZD2016008).}
}
\author{
\small{Minjia Shi$^{1,2}$, Rongsheng Wu$^1$, Liqin Qian$^1$, Lin Sok$^{1,3}$, Patrick Sol\'e$^4$}\\
\and \small{${}^1$School of Mathematical Sciences, Anhui University, Hefei, 230601, China}\\
\small{${}^2$National Mobile Communications Research Laboratory,}\\
 \small{Southeast University, 210096, Nanjing, China }\\
 \small{${}^3$Department of Mathematics, Royal University of Phnom Penh, Cambodia}\\
 \small{${}^4$CNRS/LAGA, University of Paris 8, 2 rue de la Libert\'e, 93 526 Saint-Denis, France}
}
\date{}
\maketitle
{\bf Abstract:} {In this paper, several classes of three-weight codes and two-weight codes for the homogeneous metric over the chain ring
$R=\mathbb{F}_p+u\mathbb{F}_p+\cdots +u^{k-1}\mathbb{F}_{p},$ with $u^k=0,$ are constructed, which generalises \cite{SL}, the special case of $p=k=2.$
These codes are defined as trace codes. In some cases of their defining sets, they are abelian. Their homogeneous weight distributions are computed by using exponential sums.
In particular, in the two-weight case, we give some conditions of optimality of their Gray images by using the Griesmer bound.
Their dual homogeneous distance is also given. The codewords of these codes are shown to be minimal for inclusion of supports, a fact favorable to an application to secret sharing schemes.
}


\section{Introduction}

\hspace*{0.6cm}Two-weight codes and three-weight codes form a class of combinatorial codes which are closely related to combinatorial designs, finite geometry and graph theory.
Information on them can be found in \cite{RWM,D}. Some interesting two-weight and three-weight codes were presented in \cite{DLC,DD,ZHQ,HZL}. It is worth mentioning that part of the codes they obtained have new parameters and nice access structures. A topical application of few weights
codes is their use in the  Massey scheme \cite{MM} for secret sharing, an important topic in cryptography and computer security. In that application the poset based on the codewords ordered by inclusion
of support plays a crucial role \cite{AB,YC}.

Recently, the authors have constructed several infinite families of binary and $p$-ary few weights codes from trace codes over $\mathbb{F}_2+u\mathbb{F}_2$, $\mathbb{F}_p+u\mathbb{F}_p$, and a non-chain ring, respectively in \cite{SL,SWY,SW}. In addition, the authors have investigated the minimal codewords of the codes they constructed, and determined all the codewords are minimal.

In the present paper, following this trend, we use as alphabet the larger ring $R=\mathbb{F}_p+u\mathbb{F}_p+\cdots +u^{k-1}\mathbb{F}_p,$
where $p$ is a prime number and $u^k=0.$ Although most of previous work on two-weight and three-weight codes were done on cyclic codes and cyclotomy \cite{AE}, the first two families
of codes
we construct here are provably abelian. Their homogeneous weight distributions are determined by using exponential sums, Gauss sums in the second family, and sums similar
to those in \cite{ZHQ,HZL} for the third family. By a Gray map, we obtain several infinite families of $p$-ary
two-weight and three-weight codes. As for the two-weight case, the image codes are shown to be optimal for given length and dimension by the application of the Griesmer bound under some conditions \cite{GG}.

The manuscript is organized as follows. Section 2 fixes some notations and definitions for this paper.
Section 3 presents the main results. The optimality and the dual homogeneous distance are discussed in Section 4. Section 5
determines the support structure of their Gray images, and the application to secret sharing schemes is given.
Section 6 summarizes this paper and gives some challenging open problems.

\section{Preliminaries}
\subsection{The ring extension of $R$}
\hspace*{0.6cm}Let $\mathcal{R}=\mathbb{F}_{p^m}+u\mathbb{F}_{p^m}+\cdots +u^{k-1}\mathbb{F}_{p^m}$, which is a ring extension of $R$ of degree $m$, and $m$ is a positive integer.
There is a generalized {\bf Trace function}, denoted by $Tr$, from $\mathcal{R}$ down to $R,$ and defined as $Tr(a_0+a_1u+\cdots +a_{k-1}u^{k-1})=tr(a_0)+tr(a_1)u+\cdots +tr(a_{k-1})u^{k-1},$
 for all $a_i\in \mathbb{F}_{p^m}$ and $i=0,1,\ldots ,k-1$.
Here $tr()$ denotes the standard trace of $\mathbb{F}_{p^m}$ down to $\mathbb{F}_{p}.$

\subsection{Gray map}
 \hspace*{0.6cm}Any integer $z$ can be written uniquely in base $p$ as $z=p_0(z)+pp_1(z)+p^2p_2(z)+\cdots$, where $0 \leq p_i(z)\leq p-1, i=0,1,2,\ldots$. The {\bf Gray map} $\Phi :R \rightarrow \mathbb{F}_p^{p^{k-1}}$ is defined as follows:
 $$\Phi(a)=(b_0,b_1,b_2,\ldots,b_{p^{k-1}-1}),$$
 where $a=a_0+a_1u+\cdots+a_{k-1}u^{k-1}$. Then for all $0\leq i\leq p^{k-2}-1,0\leq \epsilon \leq p-1,$ we have
 \begin{equation*}
   \small
 b_{ip+\epsilon}=\begin{cases}
 \emph{ }a_{k-1}+\sum \limits_{l=1}^{k-2}p_{l-1}(i)a_l+\epsilon a_0,\emph{ }\emph{ }\emph{ }\emph{ }\emph{ }\emph{ }\emph{ }$if$\  k\geq 3, \notag  \\
   \emph{ }a_1+\epsilon a_0,\emph{ }\emph{ }\emph{ }\emph{ }\emph{ }\emph{ }\emph{ }\emph{ }\emph{ }\emph{ }\emph{ }\emph{ }\emph{ }\emph{ }\emph{ }\emph{ }\emph{ }\emph{ }\emph{ }\emph{ }\emph{ }\emph{ }\emph{ }\emph{ }\emph{ }\emph{ }\emph{ }\emph{ }$if$ ~ k=2.\notag \\
\end{cases}
 \end{equation*}
 For instance, when $p=k=2$, it is easy to check that the Gray map adopted in the trace codes of \cite{SL} is the same as the Gray map defined here. As an additional example, when $p=k=3$, write $\Phi(a_0+a_1u+a_2u^2)=(b_0,b_1,b_2,\ldots,b_8)$.
 According to the definition above, we have $0\leq i\leq 2$, $0\leq \epsilon\leq 2$ and $\sum \limits_{l=1}^{k-2}p_{l-1}(i)a_l=p_0(i)a_1=ia_1$. Then we get
 $$b_0=a_2,b_1=a_2+a_0,b_2=a_2+2a_0,b_3=a_2+a_1,b_4=a_2+a_1+a_0,$$
 $$b_5=a_2+a_1+2a_0,b_6=a_2+2a_1,b_7=a_2+2a_1+a_0,b_8=a_2+2a_1+2a_0.$$
  It is easy to extend the Gray map from $R^n$ to $\mathbb{F}_p^{p^{k-1}n}$, and we also know from \cite{SS} that $\Phi$ is injective and linear.
\subsection{Homogeneous metric}
\hspace*{0.6cm}For $x=(x_1,x_2,\ldots,x_n)$ and $y=(y_1,y_2,\ldots,y_n)\in \mathbb{F}_p^n, \ d_H(x,y)=|\{i:x_i\neq y_i\}|$ is called the Hamming distance between $x$ and $y$ and $w_H(x)=d_H(x,0)$, the Hamming weight of $x$. The Hamming weight of a codeword $c=(c_1,c_2,\ldots,c_n)$ of $\mathbb{F}_p^n$ can also be equivalently defined as $w_H(c)=\sum\limits_ {i=1}^{n}w_H(c_i)$, where $w_H(c_i)$ equals 0
if and only if $c_i$ is a zero element.

The {\bf homogeneous weight} of an element $x\in R$ is defined as follows:
 \begin{equation*}
 \small
 w_{hom}(x)=\begin{cases}
 \emph{ }0,\emph{ }\emph{ }\emph{ }\emph{ }\emph{ }\emph{ }\emph{ }\emph{ }\emph{ }\emph{ }\emph{ }\emph{ }\emph{ }\emph{ }\emph{ }\emph{ }\emph{ }\emph{ }$if$\emph{ } x=0, \notag  \\
   \emph{ }p^{k-1},\emph{ }\emph{ }\emph{ }\emph{ }\emph{ }\emph{ }\emph{ }\emph{ }\emph{ }\emph{ }\emph{ }\emph{ }\emph{ }\emph{ }$if$\emph{ } x\in ( u^{k-1})\backslash \{0\},\notag \\
  \emph{ } (p-1)p^{k-2}, \emph{ }\emph{ }\emph{ }\emph{ }\emph{ }\emph{ }$if$\emph{ } x\in R\backslash ( u^{k-1}). \notag\\
\end{cases}
\end{equation*}

The homogeneous weight of a codeword $c=(c_1,c_2,\ldots,c_n)$ of $R^n$ is defined as $w_{hom}(c)=\sum\limits_{i=1}^nw_{hom}(c_i)$. For any $x,y\in R,$ the {\bf homogeneous distance} $d_{hom}$ is given by $d_{hom}(x,y)=w_{hom}(x-y)$. As was observed in \cite{SS}, $\Phi$ is a distance preserving isometry from $(R^n,d_{hom})$ to $(\mathbb{F}_p^{p^{k-1}n},d_H)$, where $d_{hom}$ and $d_H$ denote the homogeneous and Hamming distance in $R^n$ and $\mathbb{F}_p^{p^{k-1}n}$, respectively. This means if $C$ is a linear code over $R$ with parameters $(n,p^t,d)$, then $\Phi(C)$ is a linear code of parameters $[p^{k-1}n,t,d]$ over $\mathbb{F}_p$. Note that when $p=k=2$, the homogeneous weight is none other than the Lee weight, which was considered in \cite{SL}.

\subsection{Character sums}
\hspace*{0.6cm}Throughout this paper, let $q=p^m$. Now we present some basic facts about Gauss sums. Denote the canonical additive characters of $\F_p$ and $\F_q$ by $\phi,\chi$, respectively. Denote the multiplicative characters of $\F_p$ and $\F_q$ by $\lambda,\psi$, respectively. The \textbf{Gauss sums} over $\F_p$ and $\F_q$ are defined respectively by
$$G(\lambda,\phi)=\sum_{x\in \F_p^*}\lambda(x)\phi(x),~~~G(\psi,\chi)=\sum_{x\in \F_{q}^*}\psi(x)\chi(x).$$

Assume $q$ is odd and let $\eta$ be a quadratic multiplicative character of $\mathbb{F}_q$, which is defined by $\eta(x)=1,$ if $x$ is the square of an element of $\mathbb{F}_q^\ast$ and $\eta(x)=-1$ otherwise. Then, we define the following character sums
\begin{eqnarray*}
\overline{Q}=\sum_{x\in {\mathcal{Q}}}\chi(x), \ \overline{N}=\sum_{x\in {\mathcal{N}}}\chi(x),
\end{eqnarray*}
where $\mathcal{Q}$ denotes the set of squares in $\mathbb{F}_q$ and $\mathcal{N}$ denotes the set of nonsquares in $\mathbb{F}_q$. By orthogonality of characters \cite[Lemma 9]{MJ}, it is easy to check that $\overline{Q}+\overline{N}=-1.$
Noting that the characteristic function of ${\mathcal{Q}}$ is $\frac{1+\eta}{2},$ then we get
\begin{eqnarray*}
\overline{Q}=\frac{G(\eta)-1}{2}, \ \overline{N}=\frac{-G(\eta)-1}{2}.
\end{eqnarray*}
Let $(\frac{a}{p})$ denote the Legendre symbol for a prime $p$ and an integer $a$. The quadratic Gauss sums are  well known \cite{DY}, and given as follows:
$$G(\eta)=(-1)^{m-1}\sqrt{(p^\ast)^m},$$
where $p^\ast=(\frac{-1}{p})p=(-1)^{\frac{p-1}{2}}p.$

\subsection{Trace codes with defining sets}
\hspace*{0.6cm}Let $\mathcal{D}$ be a subset of $\mathcal{R}^*$, then we define a linear code over $R$ as follows:
$$C_{\mathcal{D}}=\{(Tr(xd))_{d\in \mathcal{D}}:x\in \mathcal{R} \}.$$
$\mathcal{D}$ is called the defining set of the code $C_\mathcal{D}$. The selection of $\mathcal{D}$ directly affects the constructed linear code, we can obtain few weights codes by the proper selection of $\mathcal{D}$. In this subsection, we will give three defining sets of $C_\mathcal{D}$, and the weight enumerator is computed in the next section.
\begin{itemize}
\item \textbf{The first definition set $\mathcal{D}_1$:}\\
$\mathcal{D}_1=\mathcal{Q}\times \mathbb{F}_{p^m}\times \cdots \times \mathbb{F}_{p^m}$, so that $|\mathcal{D}_1|=\frac{p^{(k-1)m}(p^m-1)}{2}$, where $\mathcal{Q}$ denotes the set of squares in $\mathbb{F}_{p^m}$.
\end{itemize}
\begin{itemize}
\item \textbf{The second definition set $\mathcal{D}_2$:}\\
 $\mathcal{D}_2=\mathcal{R}^*$, where ${\mathcal{R}}^*$ denotes the group of units in ${\mathcal{R}}$, i.e., ${\mathcal{R}}^*=\{a_0+a_1u+\cdots +a_{k-1}u^{k-1}:a_0\in \mathbb{F}_{p^m}^{*},a_i\in \mathbb{F}_{p^m},i=1,2,\ldots ,k-1\},$ and it is immediate to check that the order of $\mathcal{R}^\ast$ is $p^{(k-1)m}(p^m-1)$. We know that the defining set $\mathcal{D}_1$ is a subgroup of $\mathcal{D}_2$ of index 2.
\end{itemize}

Before defining the third defining set, we first introduce some notations. Let $N'$ be a positive integer such that $N'|(p^m-1)$, $N'_1={\rm lcm}(N', \frac{p^m-1}{p-1})$ and $N'_2={\rm gcd}(N', \frac{p^m-1}{p-1})$. Let $\alpha$ be a fixed primitive element of $\F_{p^m}$ and $\F_{p^m}^*=\langle\alpha\rangle$.  Define $C_i^{N'}=\alpha^i\langle\alpha^{N'} \rangle, i=0,1,\cdots,N'-1,$ and $\langle\alpha^{N'}\rangle$ is a subgroup of $\F_{p^m}^*$. Let
$n_1= N'_1/N'.$ Now we can define the third defining set.
\begin{itemize}
\item \textbf{The third definition set $\mathcal{D}_3$:}
$$\mathcal{D}_3=D'+u\mathbb{F}_{p^m}+\cdots +u^{k-1}\mathbb{F}_{p^m}\subseteq \mathcal{R}^*,$$
\end{itemize}
where $D'=\{d_j=\alpha^{N'(j-1)}:j=1,2,\ldots,n_1\}\subseteq C_0^{N'}\subseteq \F_{p^m}$. Here $\{d_1, d_2, \ldots, d_{n_1}\}$ forms a complete set of coset representatives of the factor group $C_0^{N'_2}/\F_p^*$. For more details about the construction of $D'$, the reader may refer to \cite{HZL}.

\section{The main results}
\hspace*{0.6cm}In order to obtain our main results, we first introduce the following notations:
\begin{itemize}
  \item $M$ is the maximal ideal of $\mathcal{R},$ i.e., $M=(u)=\{a_1u+a_2u^2+\cdots +a_{k-1}u^{k-1}:a_i\in \mathbb{F}_{p^m},i=1,2,\ldots ,k-1\}$. 
  \item $Ev_i(a)=(Tr(ax))_{x\in \mathcal{D}_i}$, where $a$ is an element of the ring $\mathcal{R}$, and $i\in \{1,2,3\}$. $Ev_i()$ denote evaluation maps.
  \item $N_1=p^{k-1}|\mathcal{D}_1|=(p^m-1)p^{(k-1)(m+1)}/2$, $N_2=p^{k-1}|\mathcal{D}_2|=(p^m-1)p^{(k-1)(m+1)}$, and $N_3=p^{k-1}|\mathcal{D}_3|=n_1p^{(k-1)(m+1)}=\frac{N'_1}{N'}p^{(k-1)(m+1)}$.
  \item $\Re(\Delta)$ is the real part of the complex number $\Delta$.
\end{itemize}

 Abelian codes are a natural generalization of cyclic codes. Denote the ring of integers modulo $m$ by $\mathbb{Z}_{m}$. With the integer $n=s_1\cdot s_2\cdot\cdots\cdot s_r$,
 we associate the group $G=\mathbb{Z}_{s_1}\times \mathbb{Z}_{s_2}\times \cdots \times \mathbb{Z}_{s_r}.$
 An {\bf Abelian code} of length $n$ over $R$ attached to the group $G$ is an ideal in the algebra
 $$R[X_1,X_2,\ldots,X_r]/(X_1^{s_1}-1,X_2^{s_2}-1,\ldots,X_r^{s_r}-1).$$
 In other words, the code $\mathcal{C}$ over $R$ is an ideal of the group ring $R[G],$ i.e.,
 the coordinates of $\mathcal{C}$ are indexed by elements of $G$ and $G$ acts regularly on this set. In the special case when $G$ is cyclic, that is $r=1,$ the code is a cyclic code in the usual sense \cite{MJ}.
\begin{proposition} The subgroup $\mathcal{D}_1$ of $\mathcal{R}^*$ acts regularly on the coordinates of $C_{\mathcal{D}_1}.$
\begin{proof}
For any $v',u' \in \mathcal{D}_1$ the change of variables $ x\mapsto (u'/v')x$ permutes the coordinates of $C_{\mathcal{D}_1},$ and maps $v'$ to $u'.$ Such a permutation is unique, given $v',u'.$
\end{proof}
\end{proposition}

The code $C_{\mathcal{D}_1}$ is thus an {\em Abelian code} with respect to the group $\mathcal{D}_1.$ In other words, it is an ideal of the group ring $R[\mathcal{D}_1].$ As observed above, we know $C_{\mathcal{D}_1}$ may be not cyclic since $\mathcal{D}_1$ is not a cyclic group. For the defining set $\mathcal{D}_2$, the code $C_{\mathcal{D}_2}$ has
a similar property, so we will not repeat it here. The next proposition shows that the Gray images of $C_{\mathcal{D}_1}$ and $C_{\mathcal{D}_2}$ are invariant under a transitive group of permutations.
\begin{proposition}
A finite group of size $N_1$ (resp. $N_2$) acts transitively on the coordinates of $\Phi(C_{\mathcal{D}_1})$ (resp. $\Phi(C_{\mathcal{D}_2}$) ).
\begin{proof}
We just consider the case $k\geq 3$ here. Let $\textbf{a}=\textbf{a}_0+\textbf{a}_1u+\cdots +\textbf{a}_{k-1}u^{k-1}$ be a codeword of $C_{\mathcal{D}_1}$,
where $\textbf{a}_s\in \mathbb{F}_p^{N_1}$,  $s=0,1,\ldots ,k-1$. According to the definition of the Gray map, we assume that the value of the $b_{ip+\epsilon}$-th position
of the codeword $\Phi(\textbf{a})$ is $\textbf{a}_{k-1}+\sum \limits_{l=1}^{k-2}p_{l-1}(i)\textbf{a}_l+\epsilon \textbf{a}_0$, where $0\leq i\leq p^{k-2}-1$ and $0\leq \epsilon \leq p-1$. Then we consider the codewords $\textbf{d}=(1+d_1u+\cdots +d_{k-1}u^{k-1})\textbf{a}$,
where $d_j\in \mathbb{F}_p,$ $j=1,2 \ldots,k-1$, now the values of the $b_{ip+\epsilon}$-th positions of the codewords $\Phi(\textbf{d})$ are
$\textbf{a}_{k-1}+d_1\textbf{a}_{k-2}+\cdots +d_{k-1}\textbf{a}_0+\sum \limits_{l=1}^{k-2}p_{l-1}(i)\textbf{a}_l+\epsilon \textbf{a}_0$, it is easy to
check that the values of the $b_{ip+\epsilon}$-th positions of the codewords $\Phi(\textbf{d})$ run through all the values of the components in $\Phi(\textbf{a})$,
so that $\Phi(C_{\mathcal{D}_1})$ is invariant under the involution that permutes the $p^{k-1}$ parts of a codeword. We have a similar proof for $\Phi(C_{\mathcal{D}_2})$, and omit details here.
\end{proof}
\end{proposition}
Let $\omega=\exp(\frac{2\pi i}{p})$ and  $y=(y_1,y_2,\ldots,y_N)\in \mathbb{F}_p^N,$ then we define $\Theta(y)=\sum\limits_{j=1}^N\omega^{y_j}.$
For convenience, we write $\theta_i(a)=\Theta(\Phi(Ev_i(a)))$, and it can be verified that $\theta_i(sa)=\Theta(\Phi(Ev_i(sa)))$, where $i\in\{1,2,3\}$ for any $s\in \F_p^*.$ Before computing the homogeneous weight enumerator, we first state some auxiliary lemmas.
\begin{lemma}\cite[Lemma 1]{SW} For all $y=(y_1,y_2,\ldots,y_N)\in \mathbb{F}_p^N,$ we have
$$\sum_{s=1}^{p-1}\Theta(sy)=(p-1)N-pw_H(y).$$
\end{lemma}
According to Lemma 1, we can check that for any codeword $Ev_i(a)$ of $C_{\mathcal{D}_i}$, $i\in\{1,2,3\}$, we have
$$w_{hom}(Ev_i(a))=\frac{(p-1)N-\sum\limits_{s=1}^{p-1}\theta_i(sa)}{p}.$$

The following lemma is the key to the study of the Gray images $\Phi(C_{\mathcal{D}_i})$, $i\in \{1,2,3\}$, and it guarantees that the dimension of the image code is $km$. The trace function is nondegenerate here, and the proof is easy, so we omit it.
\begin{lemma}
Fix $i\in \{1,2,3\}.$ If for some $a,b\in \mathcal{R}$  and all $x \in\mathcal{D}_i,$ we have $Tr(ax)=Tr(bx),$ then $a=b.$
\end{lemma}
Now, we discuss the homogeneous weight of the codewords in $C_{\mathcal{D}_1}$ based on two cases. If $m$ is even and $p$ is odd prime, we will get an infinite class of three-weight codes, while we will obtain an infinite class of two-weight codes when $m$ is odd and $p\equiv 3 \pmod{4}$.
\subsection{The first defining set $\mathcal{D}_1$}
\subsubsection{$m$ is even}
\hspace*{0.6cm}The following lemma is important to simplify the proof of case (b) in Theorem 1.
\begin{lemma} Let $a=a_1u+a_2u^2+\cdots +a_{k-1}u^{k-1}\in M\backslash \{0\}$, $x=x_0+x_1u+\cdots +x_{k-1}u^{k-1}\in \mathcal{D}_1$ and $B=\sum\limits_{i=1}^{k-2}a_ix_{k-1-i}$. Then
$\sum\limits_{x_1,\ldots,x_{k-2} \in \mathbb{F}_{p^m}}\omega^{tr(B)}\neq 0$ if and only if $a_i=0$ for $i=1,2,\ldots ,k-2$. Furthermore, we have $\sum\limits_{x_1,\ldots,x_{k-2} \in \mathbb{F}_{p^m}}\omega^{tr(B)}=p^{(k-2)m}$ when $a_1=a_2=\cdots =a_{k-2}=0.$
\begin{proof}
 Suppose otherwise that there exists an $a_j\neq 0$, for $j\in \{1,2,\ldots,k-2\}$, such that $\sum\limits_{x_1,\ldots,x_{k-2} \in \mathbb{F}_{p^m}}\omega^{tr(B)}\neq 0$, then we just need to consider the term $\sum \limits_{x_{k-1-j}\in \mathbb{F}_{p^m}}\omega^{tr(a_jx_{k-1-j})}$, which equals to zero, so $\sum \limits_{x_1,\ldots,x_{k-2} \in \mathbb{F}_{p^m}}\omega^{tr(B)}= 0$, a contradiction. If $a_i=0$ for $i=1,2,\ldots ,k-2$, it is easy to check that $\sum \limits_{x_1,\ldots,x_{k-2} \in \mathbb{F}_{p^m}}\omega^{tr(B)}=p^{(k-2)m}$. The proof is completed.
\end{proof}
\end{lemma}
\begin{theorem}Assume $a\in \mathcal{R}$, if $m$ is even and $p\equiv 1 \pmod{4}$, then the homogeneous weight distribution of the codewords in $C_{\mathcal{D}_1}$ is given below.
\begin{enumerate}
{\item[(a)] If $a=0$, then $w_{hom}(Ev_1(a))=0$;
\item[(b)] If $a\in M\backslash \{0\}$, then\\
if $a\in M\backslash \{a_{k-1}u^{k-1}: a_{k-1} \in \mathbb{F}_{p^m}\}$,  then $w_{hom}(Ev_1(a))=\frac{p-1}{p}N_1$,\\
if $a=a'_{k-1} u^{k-1}$, where $a'_{k-1}\in \mathbb{F}_{p^m}^{*}$, then if\\
  \hspace*{0.6cm}$a'_{k-1}\in \mathcal{Q}$, then $w_{hom}(Ev_1(a))=\frac{p-1}{p}\Big(N_1+p^{(k-1)(m+1)}(p^{\frac{m}{2}}+1)/2\Big), $ \\
  \hspace*{0.6cm}$a'_{k-1}\in \mathcal{N}$, then $w_{hom}(Ev_1(a))=\frac{p-1}{p}\Big(N_1-p^{(k-1)(m+1)}(p^{\frac{m}{2}}-1)/2\Big); $
\item[(c)] If $a\in \mathcal{R}^*$, then $w_{hom}(Ev_1(a))=\frac{p-1}{p}N_1$.
}
 \end{enumerate}
\begin{proof}
Since $m$ is even, it is easy to verify that $s\in \mathbb{F}_p^*$ is always a square in $\mathbb{F}_{p^m}.$ Thus $\theta_1(sa)=\theta_1(a),$ for any $s\in \mathbb{F}_p^*.$ Let $x=x_0+x_1u+\cdots +x_{k-1}u^{k-1}\in \mathcal{D}_1$, where $x_0\in \mathcal{Q}$ and $x_i\in \mathbb{F}_{p^m}$ for $i=1,2,\ldots ,k-1$.

(a) If $a=0$, then $Ev_1(a)=(\underbrace{0,0,\cdots,0}\limits_{|\mathcal{D}_1|})$. So $w_{hom}(Ev_1(a))=0$.

(b) Let $a=a_1u+a_2u^2+\cdots +a_{k-1}u^{k-1}\in M\backslash \{0\}$, by a direct calculation we get
\begin{eqnarray*}
  Tr(ax) &=& tr(a_1x_0)u+tr(a_1x_1+a_2x_0)u^2+\cdots +tr(a_1x_{k-2}+\cdots +a_{k-1}x_0)u^{k-1} \\
&=:&B_0+B_1u+B_2u^2+\cdots +B_{k-1}u^{k-1}.
\end{eqnarray*}

Let $I=\{I_0,I_1,I_2,\ldots ,I_{k-2}\}$, where $I_t=\{B_t, 2B_t,\ldots,(p-1)B_t\}$, $0\leq t\leq k-2$, so we know that $I$ is a set with $(p-1)(k-1)$ elements.
According to the Gray map defined in Subsection 2.2, we can write $\Phi(Tr(ax))=(A_0,A_1,A_2,\ldots,A_{k-1})$, where $A_0=B_{k-1}$, $A_j=\{B_{k-1}+b_{i_1}+b_{i_2}+\cdots +b_{i_j}|b_{i_f\in I_t},\ 1\leq f \leq j \leq k-1\}$ and $b_{i_f}$ are in the different sets $I_t$.
Therefore, we have
\begin{eqnarray*}
  \Phi(Ev_1(a)) &=& (B_{k-1},A_1,A_2,\ldots ,A_{k-1})_{x_0,\ldots ,x_{k-1}}.
\end{eqnarray*}
Since each component of $\Phi(Ev_1(a))$ contains $B_{k-1}$, using Lemma 3, it is easy to know that $\theta_1(a)=\Theta(\Phi(Ev_1(a)))=0$ if and only if $a\in M\backslash \{a_{k-1} u^{k-1}\}$, where $a_{k-1} \in \mathbb{F}_{p^m}$, which implies $w_{hom}(Ev_1(a))=\frac{p-1}{p}N_1$ by the application of Lemma 1.

If $a\overline{\in} M\backslash \{a_{k-1} u^{k-1}\}$, i.e., $a=a'_{k-1}u^{k-1}$, where $a'_{k-1}\in \mathbb{F}_{p^m}^\ast$, then we have $ax=a'_{k-1}x_0u^{k-1}$. Thus
$$Tr(ax)=tr(a'_{k-1}x_0)u^{k-1}=:Du^{k-1},$$
and then
$$\Phi(Ev_1(a))=(\underbrace{D,D,\ldots ,D} \limits_{p^{k-1}})_{x_0,x_1,\ldots ,x_{k-1}}.$$
This gives
\begin{eqnarray*}
  \theta_1(a) &=& \Theta(\Phi(Ev_1(a)))=p^{k-1}\sum_{x_0\in \mathcal{Q}}\  \ \sum_{x_1,\ldots,x_{k-1}\in \mathbb{F}_{p^m}}\omega^D \\
            & =& p^{k-1} p^{(k-1)m}\sum_{x_0\in \mathcal{Q}}\omega^D.
\end{eqnarray*}
 After variable substitution, we see that the term $\sum\limits_{x_0\in \mathcal{Q}}\omega^D$ equals $\overline{Q}$ or $\overline{N}$ depending on $a'_{k-1} \in {\mathcal{Q}}$ or $a'_{k-1} \in {\mathcal{N}}.$ Because $m$ is even and $p\equiv 1 \pmod{4}$, $G(\eta)=-p^\frac{m}{2},\ \overline{Q}=(-p^\frac{m}{2}-1)/2$ and $\overline{N}=(p^\frac{m}{2}-1)/2$. Then the statement follows from Lemma 1, i.e., if $a'_{k-1} \in \mathcal{Q},$ $w_{hom}(Ev_1(a))=\frac{p-1}{p}(N_1-p^{(k-1)(m+1)}\overline{Q})$ or $w_{hom}(Ev_1(a))=\frac{p-1}{p}(N_1-p^{(k-1)(m+1)}\overline{N})$ when $a'_{k-1}\in \mathcal{N}$.

(c) Let $a=a_0+a_1u+\cdots +a_{k-1}u^{k-1}\in \mathcal{R}^*$, by a direct calculation we have
\begin{eqnarray*}
 Tr(ax) &=&tr(a_0x_0)+tr(a_0x_1+a_1x_0)u+\cdots+tr(a_0x_{k-1}+a_1x_{k-2} \\
 &&+\cdots +a_{k-1}x_0)u^{k-1}\\
 &=:&E_0+E_1u+\cdots +E_{k-1}u^{k-1}.
\end{eqnarray*}
Since $\sum\limits_{x_{k-1}\in \mathbb{F}_{p^m}}\omega^{a_0x_{k-1}}=0$, we can easily check that
 $$ \sum_{x_0\in \mathcal{Q}} \ \sum_{x_1,x_2,\ldots ,x_{k-1}\in \mathbb{F}_{p^m}} \ \omega^{E_{k-1}}=0.$$
 Note that each component of the Gray image $\Phi(Ev_1(a))$ contains $E_{k-1}$, so we can get $\theta_1(a)=0.$ Following Lemma 1, we obtain $w_{hom}(Ev_1(a))=\frac{p-1}{p}N_1$.
\end{proof}
\end{theorem}
\begin{remark}
Theorem 1 together with Lemma 2 imply $\Phi(C_{\mathcal{D}_1})$ is a $p$-ary code of length $N_1$, dimension $km$, with three nonzero weights $w_1<w_2<w_3$ of values
\begin{eqnarray*}
  w_1 &=&\frac{p-1}{p}\Big(N_1-p^{(k-1)(m+1)}(p^{\frac{m}{2}}-1)/2\Big),\\
  w_2&=&\frac{p-1}{p}N_1,\\
   w_3&=&\frac{p-1}{p}\Big(N_1+p^{(k-1)(m+1)}(p^{\frac{m}{2}}+1)/2\Big),
\end{eqnarray*}
with respective frequencies $f_1,f_2,f_3$ given by
\begin{eqnarray*}
  f_1 =\frac{p^m-1}{2}, \ f_2 =p^{km}-p^m, \ f_3 =\frac{p^m-1}{2}.
\end{eqnarray*}

In the case of $p\equiv 3 \pmod{4}$, we know that $G(\eta)=p^{\frac{m}{2}}$ when $m$ is singly-even, and that $G(\eta)=-p^{\frac{m}{2}}$ when $m$ is doubly-even. We can also obtain a $p$-ary linear code with three nonzero weights by using a similar approach, we omit the proof here. It is easy to check that whether $p\equiv 1 \pmod{4}$ or $p\equiv 3 \pmod{4}$ when $m$ is even, the weight distribution of $C_{\mathcal{D}_1}$ is the same.
\end{remark}

\subsubsection{$m$ is odd and $p\equiv 3 \pmod{4}$}
\hspace*{0.6cm}In this case, we know from Subsection 2.4 that $G(\eta)$ is imaginary, i.e., $\Re(\overline{Q})=\Re(\overline{N})=-\frac{1}{2}.$ Then, we give the following correlation lemma, which establishes a linkage between $\theta(sa)$ and $\Re{(\theta(a))}$. We use a similar method in Theorem 1 to discuss the homogeneous weight distribution of $C_{\mathcal{D}_1}$.

\begin{lemma}\cite[Lemma 2]{SW}
If $p\equiv 3 \pmod{4},$ then $\sum\limits_{s=1}^{p-1}\theta(sa)=(p-1)\Re(\theta(a)).$
\end{lemma}
\begin{theorem}Assume $a\in \mathcal{R}$, if $m$ is odd and $p\equiv 3 \pmod{4},$ then the homogeneous weight distribution of the codewords in $C_{\mathcal{D}_1}$ is given below.
\begin{enumerate}
\item[(a)] If $a=0$, then $w_{hom}(Ev_1(a))=0$;
\item[(b)]  If $a\in  M\backslash \{0\}$, then\\
if $a\in M\backslash \{a_{k-1} u^{k-1}: a_{k-1} \in \mathbb{F}_{p^m}\}$,  then $w_{hom}(Ev_1(a))=\frac{p-1}{p}N_1$,\\
if $a=a'_{k-1} u^{k-1}$, where $a'_{k-1}\in \mathbb{F}_{p^m}^{*}$, then $w_{hom}(Ev_1(a))=\frac{p-1}{p}\Big(N_1+p^{(k-1)(m+1)}/2\Big)$;
\item[(c)] If $a\in \mathcal{R}^*$, then $w_{hom}(Ev_1(a))=\frac{p-1}{p}N_1$.
 \end{enumerate}
\begin{proof}
 We just give the proof of the case (b) here, the rest cases are similar to those in Theorem 1. Note that $\Re(\theta_1(a))=0$ when $a\in M\backslash \{a_{k-1} u^{k-1}:a_{k-1} \in \mathbb{F}_{p^m}\}$, and $\Re(\theta_1(a))=-\frac{p^{(k-1)(m+1)}}{2}$ when $a=a'_{k-1} u^{k-1}$, where $a'_{k-1}\in \mathbb{F}_{p^m}^{*}$. Combining Lemmas 1 with 4, then we have
 $$pw_{hom}(Ev_1(a))=(p-1)N_1-(p-1)\Re(\theta_1(a)).$$
Then the result follows.
\end{proof}
 \end{theorem}
 \begin{remark}
Combining Theorem 2 with Lemma 2, we obtain an infinite family of $p$-ary two-weight codes of parameters $[N_1,km],$ with two nonzero weights $w'_1<w'_2$ given by
\begin{eqnarray*}
w'_1=\frac{p-1}{p}N_1, \ w'_2=\frac{p-1}{p}\Big(N_1+(p^{(k-1)(m+1)})/2\Big),
\end{eqnarray*}
with respective frequencies $f'_1,f'_2$ given by
\begin{eqnarray*}
f'_1=p^{km}-p^{m}, \ f'_2=p^m-1.
\end{eqnarray*}

It is necessary to distinguish the difference between the case when $k=2$ in the present paper and the case in \cite{SW}. Although the ring and the defining set are the same as \cite{SW}, which is not the special case of this paper, because the Gray maps are different. We list their weight distributions in Tables I and II to show the difference.
\begin{center}$\mathrm{Table \  I. }$ weight distribution of the three-weight case ($k=2$)\\
\begin{tabular}{c||c||c}
\hline
  Weight in \cite[Theorem 1]{SW} & Weight in Theorem 1  & Frequency  \\
  \hline
  0        &    0   &  1              \\
  $(p^{m}-p^{m-1})(p^m-p^{\frac{m}{2}})$        &      $\frac{(p^{m+1}-p^m)(p^m-p^{\frac{m}{2}})}{2} $     &   $\frac{p^m-1}{2}$       \\
  $(p^{m}-p^{m-1})(p^m-1)$                                                  &     $\frac{(p^{m+1}-p^m)(p^m-1)}{2} $       &     $p^{2m}-p^m$     \\
  $(p^{m}-p^{m-1})(p^m+p^{\frac{m}{2}})$         &       $\frac{(p^{m+1}-p^m)(p^m+p^{\frac{m}{2}})}{2}$ &  $\frac{p^m-1}{2}$        \\
  \hline
\end{tabular}
\end{center}
\begin{center}$\mathrm{Table \  II. }$ weight distribution of the two-weight case ($k=2$)\\
\begin{tabular}{c||c||c}
\hline
  Weight in \cite[Theorem 2]{SW} & Weight in Theorem 2  & Frequency  \\
  \hline
  0        &    0   &  1              \\
  $(p^{m}-p^{m-1})(p^m-1)$        &      $\frac{(p^{m+1}-p^m)(p^m-1)}{2}$     &   $p^{2m}-p^m$    \\
  $p^{m-1}(p^{m+1}-p^m)$          &     $\frac{p^m(p^{m+1}-p^m)}{2} $       &    $p^m-1$ \\
  \hline
\end{tabular}
\end{center}

According to Tables I and II, it is easy to see that the corresponding dimension and frequency are the same. However, the nonzero weights of the codes are different. Furthermore, we can check that the corresponding nonzero weights and lengths have constant ratio in both tables. For example, in Table II, we have
$$\frac{\frac{(p^{m+1}-p^m)(p^m-1)}{2}}{(p^{m}-p^{m-1})(p^m-1)}=\frac{\frac{p^m(p^{m+1}-p^m)}{2}}{p^{m-1}(p^{m+1}-p^m)}=\frac{p}{2},$$
and we know the length of the codes has the same proportional relationship, namely,
$$\frac{(p^m-1)p^{m+1}}{(p^{2m}-p^{m})}=\frac{p}{2}.$$
\end{remark}

\subsection{The second defining set $\mathcal{D}_2$}
\hspace*{0.6cm}In this subsection, we will discuss the homogeneous weight of the codewords in $C_{\mathcal{D}_2}$. Using the similar method in Theorem 1, we give the next theorem without proof.
\begin{theorem}Assume $p$ is a prime number and $a\in \mathcal{R}$, then the homogeneous weight distribution of the codewords in $C_{\mathcal{D}_2}$ is given below.
\begin{enumerate}
{\item[(a)] If $a=0$, then $w_{hom}(Ev_2(a))=0$;
\item[(b)] If $a\in M\backslash \{0\}$, then\\
if $a\in M\backslash \{a_{k-1} u^{k-1}: a_{k-1} \in \mathbb{F}_{p^m}\}$,  then $w_{hom}(Ev_2(a))=\frac{p-1}{p}N_2$,\\
if $a=a'_{k-1} u^{k-1}$, where $a'_{k-1} \in \mathbb{F}_{p^m}^{*}$, then $w_{hom}(Ev_2(a))=\frac{p-1}{p}\Big(N_2+p^{(k-1)(m+1)}\Big); $
\item[(c)] If $a\in \mathcal{R}^*$, then $w_{hom}(Ev_2(a))=\frac{p-1}{p}N_2$.
}
 \end{enumerate}
\end{theorem}
\begin{remark}
Theorem 3 together with Lemma 2, we can obtain $\Phi(C_{\mathcal{D}_2})$ is a $p$-ary code of length $N_2$, dimension $km$, with two nonzero weights $w''_1<w''_2$ of values
\begin{eqnarray*}
  w''_1 =\frac{p-1}{p}N_2, \ \ \ w''_2=\frac{p-1}{p}\Big(N_2+p^{(k-1)(m+1)}\Big),
\end{eqnarray*}
with respective frequencies $f''_1,f''_2$ given by
\begin{eqnarray*}
  f''_1 =p^{km}-p^m, \ f''_2 =p^m-1.
\end{eqnarray*}

Note that when $p=k=2$, the weight distribution of $C_{\mathcal{D}_2}$ is exactly the same in \cite{SL}. This means Theorem 3 includes Theorem 1 in \cite{SL} as a special case.

On the other hand, according to Theorems 2 and 3 in this section, we have obtained  two infinite classes of $p$-ary two-weight codes, it is easy to check that the corresponding dimension and frequency are the same, and the corresponding length and the nonzero weights have constant ratio, i.e.,
$$\frac{w'_1}{w''_1} = \frac{w'_2}{w''_2} =\frac{N_1}{N_2} =\frac{(p^m-1)p^{(k-1)(m+1)}/2}{(p^m-1)p^{(k-1)(m+1)}}=\frac{1}{2}, $$
where $\ w'_1$ and $w'_2$ can be found in Remark 2. However, the conditions on $p$ and $m$ in Theorems 2 and 3 are different.
\end{remark}

\subsection{The third defining set $\mathcal{D}_3$ }
\hspace*{0.6cm}Let $D'=\{d_j=\alpha^{N'(j-1)}:j=1,2,\ldots,n_1\}\subseteq \F_{p^m}$ introduced in Subsection 2.5, then a linear code over $\mathbb{F}_p$ of length $n_1$ is defined by
$$C_{D'}=\{(tr(xd_1),tr(xd_2),\ldots,tr(xd_{n_1})):x\in \mathbb{F}_{p^m}\}.$$

Before giving the parameters of the code $C_{\mathcal{D}_3}$, we first introduce some weight formulas for the code $C_{D'}$, which can be found in \cite{HZL}. Let $c_b$ be a nonzero codeword in $C_{D'}$, we write $c_b$ as $(tr(bd_1),tr(bd_2),\ldots,tr(bd_{n_1}))$, where $b\in \F_q^*$. Then we define
$$N(b)=| \{1\leq j\leq n_1:tr(bd_j)=0 \}|,$$
and thus $w_H(c_b) =n_1-N(b).$ From \cite{HZL}, we know
\begin{equation}\label{c}
  pN(b)=n_1+\frac{1}{N'_2}\sum_{j=0}^{N'_2-1}G(\bar{\varphi^j},\chi)\varphi^j(b),
\end{equation}
where $\varphi$ is a multiplicative character of order $N'_2$ in $\widehat{\F}_q^*$ and $N'_2=\rm{gcd}(N',\frac{p^m-1}{p-1})$. Here, $\widehat{\F}_q^*$ denotes multiplicative character group.

\begin{theorem}Let $N'_2=1$. Assume $m$ is even or $m$ is odd and $p\equiv 3~({\rm mod} ~4)$.
\begin{enumerate}
{\item[(a)] If $a=0$, then $w_{hom}(Ev_3(a))=0$;
\item[(b)] If $a\in M\backslash \{0\}$, then\\
   if $a\in M\backslash \{a_{k-1}u^{k-1}: a_{k-1} \in \mathbb{F}_{p^m}\}$,  then $w_{hom}(Ev_3(a))=(p^m-1)p^{(k-1)(m+1)-1}$,\\
   if $a=a'_{k-1} u^{k-1}$, where $a'_{k-1}\in \mathbb{F}_{p^m}^{*}$, then $w_{hom}(Ev_3(a))=p^{k(m+1)-2};$
\item[(c)] If $a\in \mathcal{R}^*$, then $w_{hom}(Ev_3(a))=(p^m-1)p^{(k-1)(m+1)-1}$.
}
 \end{enumerate}
\begin{proof}
It is suffices to prove the second condition in case $(b)$, the proof of the remaining cases are similar to that of Theorems 1 and 2. Let $x=x_0+x_1u+\cdots +x_{k-1}u^{k-1}\in \mathcal{D}_{3}$ and $a=a'_{k-1}u^{k-1}$, where $a'_{k-1}\in \mathbb{F}_{p^m}^{*}$, by a direct calculation we get
$$Tr(ax)=Tr(a'_{k-1}x_0u^{k-1})=tr(a'_{k-1}x_0)u^{k-1}=:Fu^{k-1}.$$
Employing the Gray map yields
$$\Phi(Ev_3(a))=(\underbrace{F, F,\ldots, F}\limits_{p^{k-1}})_{x_0, x_1, \ldots, x_{k-1}}.$$
Then we have
$$w_{hom}(Ev_3(a))=w_H(\Phi(Ev_3(a)))=p^{(k-1)(m+1)}(n_1-N(a'_{k-1})),$$
where $N(a'_{k-1})=|\{x_0\in D':tr(a'_{k-1}x_0)=0\}|.$ When $N'_2=1$, we know $n_1=\frac{p^m-1}{p-1}$, by Formula (1), we know $pN(a'_{k-1})=n_1-1$, which implies $w_{hom}(Ev_3(a))=p^{m-1}p^{(k-1)(m+1)}$.
\end{proof}
\end{theorem}

\begin{remark}
Theorem 4 together with Lemma 2 implies that $\Phi(C_{\mathcal{D}_3})$ is a $p$-ary code of length $\frac{(p^m-1)p^{(k-1)(m+1)}}{p-1}$, dimension $km$,
with two nonzero weights $w'''_1<w'''_2$ of values
\begin{eqnarray*}
  w'''_1 =(p^m-1)p^{(k-1)(m+1)-1}, \ w'''_2=p^{k(m+1)-2},
\end{eqnarray*}
with respective frequencies $f'''_1, f'''_2$ given by
\begin{eqnarray*}
  f'''_1 =p^{km}-p^m, \ f'''_2 =p^m-1.
\end{eqnarray*}

So far, we have obtained three infinite classes of $p$-ary two-weight codes in Theorems 2, 3 and 4, but they have different parameters. In Remark 3, we have compared the difference and correlation between the first two infinite classes of two-weight codes in Theorem 2 and Theorem 3, so it suffices to compare the parameters of the codes obtained from Theorems 3 and 4. For convenience, we list their weight distributions in Table III to show the difference.
\begin{center}$\mathrm{Table \  III. }$ weight distribution of $C_{\mathcal{D}_2}$ and $C_{\mathcal{D}_3}$\\
\begin{tabular}{c||c||c}
\hline
  Weight in Theorem 3 & Weight in Theorem 4  & Frequency  \\
  \hline
  0        &    0   &  1              \\
  $(p-1)(p^m-1)p^{(k-1)(m+1)-1}$        &      $(p^m-1)p^{(k-1)(m+1)-1}$     &   $p^{km}-p^m$    \\
  $(p-1)p^{k(m+1)-2}$         &     $p^{k(m+1)-2}$       &    $p^m-1$ \\
  \hline
\end{tabular}
\end{center}

Similar to the discussion in Remark 3, it is easy to see that the corresponding dimension and frequency are the same, and the nonzero weights and length have constant ratio $p-1$. However, the conditions on $p$ and $m$ in Theorems 3 and 4 are different.
\end{remark}

\begin{example}
Let $(p,m,k)=(3,3,2)$. If $N'=2$, then $N'_2=1$. In the light of Theorem $4$, we can obtain $\Phi(C_{\mathcal{D}_3})$ is a $[1053,6,702]$ ternary code, the nonzero weights are $702$ and $729$ and the corresponding frequencies are $702$ and $26$, respectively.
\end{example}

\begin{theorem}
Suppose $1<N'_2<p^{\frac{m}{2}}+1$. Assume $m$ is even or $m$ is odd and $p\equiv 3\pmod{4}$. Then $C_{\mathcal{D}_3}$ is a $(N_3,p^{km},d_{hom})$ linear code over $R$ which has at most $N'_2+1$ nonzero homogeneous weights, and
$$p^{(k-1)(m+1)-1}\cdot\frac{p^m-(N'_2-1)p^{\frac{m}{2}}}{N'_2}\leq d_{hom}\leq p^{(k-1)(m+1)-1}\cdot\frac{p^m-1}{N'_2}.$$
\begin{proof}
We also assume that $x=x_0+x_1u+\cdots +x_{k-1}u^{k-1}\in \mathcal{D}_{3}$ and $a=a'_{k-1} u^{k-1}$, where $a'_{k-1}\in \F^*_{p^m}$. We know from the proof of Theorem 4 that $w_{hom}(Ev_3(a))=w_H(\Phi(Ev_3(a)))=p^{(k-1)(m+1)}(n_1-N(a'_{k-1}))$, where $N(a'_{k-1})=|\{x_0\in D':tr(a'_{k-1}x_0)=0\}|.$ From Formula (1) and $n=\frac{p^m-1}{(p-1)N'_2}$, we have
\begin{eqnarray*}
  n-N(a'_{k-1}) &=& n-\frac{n}{p}-\frac{\sum\limits_{j=0}^{N'_2-1}G(\bar{\varphi^j},\chi)\varphi^j(a'_{k-1})}{pN'_2}\\
  &=&\frac{p^m}{pN'_2}-\frac{\sum\limits_{j=1}^{N'_2-1}G(\bar{\varphi^j},\chi)\varphi^j(a'_{k-1})}{pN'_2}.
\end{eqnarray*}
Note that $\bigg|\sum\limits_{j=1}^{N'_2-1}G(\bar{\varphi^j},\chi)\varphi^j(a'_{k-1})\bigg|\leq (N'_2-1)p^{\frac{m}{2}}.$ Since $N'_2<p^{\frac{m}{2}}+1$, then
$$p^{(k-1)(m+1)-1}\cdot\frac{p^m-(N'_2-1)p^{\frac{m}{2}}}{N'_2}\leq w_{hom}(Ev_3(a))\leq p^{(k-1)(m+1)-1}\cdot\frac{p^m+(N'_2-1)p^{\frac{m}{2}}}{N'_2},$$
here $\varphi$ is a multiplicative character of order $N'_2$ in $\widehat{\F}_q^*$, so the above case gives at most $N'_2$ nonzero homogeneous weights. The other cases give only a nonzero homogeneous weight $\frac{p-1}{p}N_3$ by a similar discussion in Theorems 1 and 2. Hence the code $C_{\mathcal{D}_3}$ has at most $N'_2+1$ nonzero weights. In addition, it is easy to check that $\frac{p-1}{p}N_3<p^{(k-1)(m+1)-1}\cdot\frac{p^m+(N'_2-1)p^{\frac{m}{2}}}{N'_2}$. This completes the proof.
\end{proof}
\end{theorem}
\begin{theorem}
Let $m$ be even and $N'_2>2$. Assume there exists a positive integer $k'$ such that $p^{k'} \equiv -1~({\rm mod} ~N'_2)$. Let $t=\frac{m}{2k'}$.
\begin{enumerate}
  \item [(a)] If $N'_2$ is even, $p,t$ and $\frac{p^{k'}+1}{N'_2}$ are odd, then the code $C_{\mathcal{D}_3}$ is a three-weight linear
  code provided that $N'_2<p^{\frac{m}{2}}+1$, and its weight distribution is given in Table IV.
      \begin{center}$\mathrm{Table \  IV. }~~~\mathrm{weight~ distribution~ of}~ C_{\mathcal{D}_3}$\\
\begin{tabular}{ccc||cc}
\hline
  Weight& &&& Frequency  \\
  \hline
  0        & &&& 1\\
  $p^{(k-1)(m+1)-1}\cdot\frac{p^m-(N'_2-1)p^{\frac{m}{2}}}{N'_2}$        &  &&&              $\frac{p^m-1}{N'_2}$\\
  $p^{(k-1)(m+1)-1}\cdot\frac{p^m-1}{N'_2}$                                                  &  &&&              $p^{km}-p^m$\\
  $p^{(k-1)(m+1)-1}\cdot\frac{p^m+p^{\frac{m}{2}}}{N'_2}$         &  &&& $\frac{(N'_2-1)(p^m-1)}{N'_2}$\\
  \hline
\end{tabular}
\end{center}
  \item [(b)] In all other cases, the code $C_{\mathcal{D}_3}$ is a three-weight linear code provided that $p^{\frac{m}{2}}+(-1)^t(N'_2-1)>0$ and its weight distribution is given in Table V.
      \begin{center}$\mathrm{Table \  V. }~~~\mathrm{weight~ distribution~ of}~ C_{\mathcal{D}_3}$\\
\begin{tabular}{ccc||cc}
\hline
  Weight& &&& Frequency  \\
  \hline
  0        & &&& 1\\
  $p^{(k-1)(m+1)-1}\cdot\frac{p^m+(-1)^t(N'_2-1)p^{\frac{m}{2}}}{N'_2}$        &  &&&              $\frac{p^m-1}{N'_2}$\\
  $p^{(k-1)(m+1)-1}\cdot\frac{p^m-1}{N'_2}$                                                  &  &&&              $p^{km}-p^m$\\
  $p^{(k-1)(m+1)-1}\cdot\frac{p^m-(-1)^tp^{\frac{m}{2}}}{N'_2}$         &  &&& $\frac{(N'_2-1)(p^m-1)}{N'_2}$\\
  \hline
\end{tabular}
\end{center}
\end{enumerate}
\begin{proof}
In case $(a)$, let $a=a'_{k-1} u^{k-1}$, where $a'_{k-1}\in \F^*_{p^m}$, we know from Theorem 5 that this case gives at most $N'_2$ nonzero weights in the following set
 $$\Big\{\frac{p^m-1}{pN'_2}-\frac{1}{pN'_2}t_s:s=0,1,\ldots,N'_2-1\Big\},$$
 where $t_s=\sum\limits_{j=0}^{N'_2-1}G(\bar{\varphi^j},\chi)\varphi^j(a'_{k-1})$. In the proof of \cite[Theorem 4.1]{HZL}, we have
\begin{equation*}
   t_s=\small
\begin{cases}
   \emph{ }-1+(N'_2-1)p^{\frac{m}{2}}\emph{ }\emph{ }\emph{ }\emph{ }\emph{ }\emph{ }\emph{ }\emph{ }$if$\ \ s=\frac{N'_2}{2}, \notag  \\
 \emph{ }-1-p^{\frac{m}{2}}\emph{ }\emph{ }\emph{ }\emph{ }\emph{ }\emph{ }\emph{ }\emph{ }\emph{ }\emph{ }\emph{ }\emph{ }\emph{ }\emph{ }\emph{ }\emph{ }\emph{ }\emph{ }  \  $otherwise$. \notag  \\
\end{cases}
 \end{equation*}
The other cases only give a nonzero weight $\frac{p-1}{p}N_3$, i.e., $p^{(k-1)(m+1)-1}\cdot\frac{p^m-1}{N'_2}$, then the result follows. The case $(b)$ can be obtained in the same way.
\end{proof}
\end{theorem}
\begin{example}
Let $(p,m,k)=(3,4,2)$. If $N'_2=4$, then $k'=1$ and $t=2$. In the light of Theorem 6, we can obtain $\Phi(C_{\mathcal{D}_3})$ is a $[2430,8,1458]$ ternary code, the nonzero weights are $1458, 1620$ and $2187$, and the corresponding frequencies are $60, \ 6480$ and $20$, respectively.
\end{example}
\begin{remark}
In Theorems 6, when $t$ is odd, then the parameters in Table IV and Table V are the same, while $t$ is even, the parameters in both tables are different.

According to Theorems 1 and 6, we have obtained three infinite classes of $p$-ary three-weight codes. If $t$ is odd, it is easy to check the length and two of the three nonzero weights have constant ratio, i.e.,
$$\frac{w_2}{p^{(k-1)(m+1)-1}\cdot\frac{p^m-1}{N'_2}}=\frac{w_3}{p^{(k-1)(m+1)-1}\cdot\frac{p^m+p^{\frac{m}{2}}}{N'_2}}=\frac{N_1}{N_3}=\frac{(p-1)N'_2}{2},$$
where $w_2=\frac{p-1}{p}N_1$ and $w_3=\frac{p-1}{p}\Big(N_1+p^{(k-1)(m+1)}(p^{\frac{m}{2}}+1)/2\Big)$ (see Theorem 1).
Since $N'_2>2$, we can check
 $$\frac{w_1}{p^{(k-1)(m+1)-1}\cdot\frac{p^m+p^{\frac{m}{2}}}{N'_2}}=\frac{N'_2(p-1)(p^m-p^{\frac{m}{2}})}{2(p^m-(N'_2-1)p^{\frac{m}{2}})}\neq \frac{(p-1)N'_2}{2},$$
where $w_1=\frac{p-1}{p}\Big(N_1-p^{(k-1)(m+1)}(p^{\frac{m}{2}}-1)/2\Big)$ (see Theorem 1). On the other hand, the corresponding frequencies are also different. When $t$ is even, we can discuss in a similar way to show that these three infinite three-weight codes are different.
\end{remark}

\begin{remark}
 As for the defining set $\mathcal{D}_3$, we have obtained several few weights $p$-ary linear codes by a linear Gray map. Although we adopt the idea about the construction
 of the defining set $\mathcal{D}_3$ in \cite{HZL}, the results we obtained are completely different.
 For instance, Theorems 3.2 and 4.1 of \cite{HZL} lead to several classes of $p$-ary one-weight and two-weight codes, while Theorems 4 and 6 produce several classes of $p$-ary two-weight and three-weight codes.
\end{remark}

\section{Further results}
\hspace*{0.6cm}In Section 3, we have shown that $C_{\mathcal{D}}$ is a two-weight code or three-weight code depending on the choice of $m$, $p$ or other conditions. Now, we continue to explore other properties about the codes we have constructed in Section 3. In this section we will study the optimality of the image codes $\Phi(C_{\mathcal{D}})$, and the dual homogeneous distance of the codes $C_{\mathcal{D}}$.
\subsection{Optimality of the image codes $\Phi(C_{\mathcal{D}})$}
\hspace*{0.6cm}If $C$ is a linear code with parameters $[n,k,d]$, and no $[n,k,d+1]$ code exists, then we call the code $C$ optimal. The next lemma introduces the Griesmer bound, which applies specifically to linear codes over finite fields.
\begin{lemma}\cite [Griesmer bound]{GG} Let $C$ be a linear $p$-ary code of parameters $[n,K,d]$, where $K\geq 1$. Then
$$\sum_{i=0}^{K-1}\bigg\lceil \frac{d}{p^i} \bigg\rceil \le n.$$
\end{lemma}

\begin{theorem}
 Assume $m$ is odd and $p\equiv 3 \pmod{4}.$ If the code $C_{\mathcal{D}_1}$ is defined as above for given length and dimension, then $\Phi(C_{\mathcal{D}_1})$ is optimal
 if $$m\geq max\Big\{k,\Big\lfloor\frac{p^{k-1}-2k+1}{2(k-1)}\Big\rfloor +1\Big\}.$$
\begin{proof}
In the light of Theorem 2, we know $\Phi(C_{\mathcal{D}_1})$ is a $[N_1,km,d]$ code, where $d=w'_1=\frac{p-1}{p}N_1$. Next we explore the condition such that
$\sum\limits_{i=0}^{km-1}\Big\lceil \frac{d+1}{p^i} \Big\rceil > N_1$ by using the the Griesmer bound. First of all, we guarantee that $(m+1)k-m-1 \leq km-1$, i.e., $m\geqslant k.$ Then we classify the range of $i$ to determine the value of $\Big\lceil \frac{d+1}{p^i} \Big\rceil$.
\begin{itemize}
\item If $0\leqslant i\leqslant (m+1)k-m-2,$ then $\Big\lceil \frac{d+1}{p^i} \Big\rceil = \frac{(p-1)(p^m-1)}{2}p^{(m+1)k-m-2-i}+1$;
\item If $(m+1)k-m-1\leqslant i\leqslant km-1,$ then $\Big\lceil \frac{d+1}{p^i} \Big\rceil =\frac{p-1}{2}\cdot p^{(m+1)k-2-i}$.
\end{itemize}
This implies that
 \begin{eqnarray*}
       \sum_{i=0}^{km-1}\Big \lceil \frac{d+1}{p^i} \Big \rceil& =& \sum_{i=0}^{(m+1)k-m-2}\Big\lceil \frac{d+1}{p^i}\Big\rceil+\sum_{i=(m+1)k-m-1}^{km-1}\Big\lceil \frac{d+1}{p^i}\Big\rceil \\
        &=&\frac{(p-1)(p^m-1)}{2}\sum_{i=0}^{(m+1)k-m-2}(p^{(m+1)k-m-2-i}+1)+\\
        && \frac{p-1}{2}\sum_{i=(m+1)k-m-1}^{km-1} p^{(m+1)k-2-i}\\
        &=&\frac{p^m-1}{2}(p^{(k-1)(m+1)}-1)+(m+1)k-m-1+\frac{p^m-p^{k-1}}{2}.
     \end{eqnarray*}
So we need
$\sum\limits_{i=0}^{km-1}\lceil \frac{d+1}{p^i} \rceil-N_1>0,$ i.e., $km+k-m-\frac{1}{2}-\frac{p^{k-1}}{2}>0$, and thus we get $m\geq \Big\lfloor\frac{p^{k-1}-2k+1}{2(k-1)}\Big\rfloor +1$. In all, we have $m\geq max\Big\{k,\Big\lfloor\frac{p^{k-1}-2k+1}{2(k-1)}\Big\rfloor +1\Big\}$.
\end{proof}
\end{theorem}
\begin{theorem}
 Assume $p$ is a prime number. Then the code $\Phi(C_{\mathcal{D}_2})$ is optimal if
  $$m\geq max\Big\{k,\Big\lfloor\frac{p^{k-1}-k}{k-1}\Big\rfloor +1\Big\}.$$
\begin{proof}
Using a similar approach in Theorem 7, we know $d=w''_1=\frac{p-1}{p}N_2$ here. We first guarantee that $(k-1)(m+1) \leq km-1$, i.e., $m\geqslant k.$ Then we have the same classification about the range of $i$ to determine the value of $\Big\lceil \frac{d+1}{p^i} \Big\rceil$.
\begin{itemize}
\item If $0\leqslant i\leqslant (k-1)(m+1)-1,$ then $\Big\lceil \frac{d+1}{p^i} \Big\rceil = (p-1)(p^m-1)p^{(k-1)(m+1)-1-i}+1$;
\item If $(k-1)(m+1)\leqslant i\leqslant km-1,$ then $\Big\lceil \frac{d+1}{p^i} \Big\rceil = (p-1)\cdot p^{(m+1)k-2-i}$.
\end{itemize}
This implies that
 \begin{eqnarray*}
       \sum_{i=0}^{km-1}\Big \lceil \frac{d+1}{p^i} \Big \rceil& =& \sum_{i=0}^{(k-1)(m+1)-1}\Big\lceil \frac{d+1}{p^i}\Big\rceil+\sum_{i=(k-1)(m+1)}^{km-1}\Big\lceil \frac{d+1}{p^i}\Big\rceil \\
        &=&(p^m-1)(p^{(k-1)(m+1)}-1)+(k-1)(m+1)+p^m-p^{k-1}.
     \end{eqnarray*}
So we need
$\sum\limits_{i=0}^{km-1}\Big\lceil \frac{d+1}{p^i} \Big\rceil-N_2>0,$ i.e., $km+k-m-p^{k-1}>0$, and thus we get $m\geq \Big\lfloor\frac{p^{k-1}-k}{k-1}\Big\rfloor +1$. In all, we have $m\geq max\Big\{k,\Big\lfloor\frac{p^{k-1}-k}{k-1}\Big\rfloor +1\Big\}$.
\end{proof}
 \end{theorem}
The next theorem is about the condition for the optimality of $\Phi(C_{\mathcal{D}_3})$, the proof is the same as Theorems 7 and 8, so we will not repeat here.
\begin{theorem}
 Assume $N'_2=1$, $m$ is even or $m$ is odd and $p\equiv 3 \pmod{4}$. Then the code $\Phi(C_{\mathcal{D}_3})$ is optimal if
 $$m\geq max\Big\{k,\Big\lfloor\frac{p^{k-1}-p(k-1)+k-2}{(p-1)(k-1)}\Big\rfloor +1\Big\}.$$
 \end{theorem}

\subsection{The dual homogeneous distance of trace codes $C_{\mathcal{D}}$}

\hspace*{0.6cm} If $x=(x_1,x_2,\ldots,x_n)$ and $y=(y_1,y_2,\ldots,y_n)$ are two elements of  $R^n$, their standard inner product is defined by $\langle x,y\rangle=\sum\limits_{i=1}^nx_iy_i$, where the operation is performed in $R$. The dual code of $C_{\mathcal{D}}$ is denoted by $C_{\mathcal{D}}^\perp$ and defined as $C_{\mathcal{D}}^\perp=\{y\in R^{|\mathcal{D}|}|\langle x,y\rangle =0, \forall x\in C_{\mathcal{D}}\}.$ In this subsection, we will compute the dual homogeneous distance of $C_{\mathcal{D}}$. A property of the trace function we need is that it is nondegenerate. The following lemma plays an important role in determining the dual homogeneous distance. The process of the proof is similar to \cite[Lemma 3]{SW}, so we omit it here.
\begin{lemma}
 For a fixed element $x\in \mathcal{R}$, if $Tr(ax)=0$ for $a \in \mathcal{R}$, then $x=0.$
\end{lemma}
\begin{theorem}
 For $m\ge 2,$ the dual homogeneous distance $d_{hom}'$ of $C_{\mathcal{D}_1}$ is $2(p-1)p^{k-2}.$
\begin{proof} First, we need to show that $C_{\mathcal{D}_1}^\perp$ does not contain a codeword whose only nonzero digit has homogeneous weight $(p-1)p^{k-2}$. If not, we assume that there is a codeword of $C_{\mathcal{D}_1}^\perp$ that has a symbol $\gamma =\gamma_0+\gamma_1u+\cdots +\gamma_{k-1}u^{k-1}\in R\backslash (u^{k-1}) $ at some $x \in \mathcal{D}_1$, so we know that there at least exists a coefficient $\gamma _j \neq 0,$ where $j\in\{0,1,\ldots ,k-2\}$. Let $a=a_0+a_1u+\cdots +a_{k-1}u^{k-1}\in \mathcal{R}$ and $x=x_0+x_1u+\cdots +x_{k-1}u^{k-1}\in \mathcal{D}_1$. Then we have $\gamma Tr(ax)=0$, which gives $k$ equations with respect to the coefficients of $u^i$. Comparing the coefficients of constant terms in $\gamma Tr(ax)=0$, we have $tr(\gamma_0a_0x_0)=0$, according to Lemma 6, we know  $\gamma_0x_0=0$, but $x_0\neq 0,$ so $\gamma_0=0$. Considering the other coefficients of $u^i$ in the equation $\gamma Tr(ax)=0$, we can obtain $\gamma_1=\gamma_2=\cdots =\gamma_{k-2}=0$, a contradiction.

Next, we prove that there exists a codeword of $C_{\mathcal{D}_1}^\perp$ that has homogeneous weight $2(p-1)p^{k-2}$. Since $2(p-1)p^{k-2}>p^{k-1}$,
we need to show that $C_{\mathcal{D}_1}^\perp$ does not contain a codeword that has only one digit of homogeneous weight $p^{k-1}$.
We can use a similar approach as above to prove it, and we omit it here. Then we assume that there exists a codeword of $C_{\mathcal{D}_1}^\perp$ which has two values $\alpha=\alpha_0+\alpha_1u+\cdots +\alpha_{k-1}u^{k-1}$ and $\beta=\beta_0+\beta_1u+\cdots +\beta_{k-1}u^{k-1}\in R\backslash (u^{k-1})$ at some $x,\ y\in \mathcal{D}_1$, where $x=x_0+x_1u+\cdots+x_{k-1}u^{k-1}$ and $y=y_0+y_1u+\cdots+y_{k-1}u^{k-1}$. Thus we have $\alpha Tr(ax)+\beta Tr(ay)=0$, i.e., $k$ equations as follows:
\begin{center}
  $\left\{
  \begin{array}{ll}
    \alpha_0x_0+\beta_0y_0=0;  \\
   \alpha_0x_1+\alpha_1x_0+\beta_0y_1+\beta_1y_0=0; \\
   \vdots\\
   \alpha_0x_{k-1}+\alpha_1x_{k-2}+\cdots +\alpha_{k-1}x_0+\beta_0y_{k-1}+\beta_1y_{k-2}+\cdots +\beta_{k-1}y_0=0.
  \end{array}
\right.$
\end{center}
We can treat it as a system of homogeneous linear equations with indeterminate elements $\alpha_i,\ \beta_j$,
where $i,j\in\{0,1,\ldots ,k-1\}$. It is easy to show that this system has nonzero solutions. Due to $x_0,\ y_0\in \mathcal{Q}$, without loss of generality, we let $ \alpha_0=x_0^{-1}\neq 0$ and $\beta_0=-y_0^{-1}\neq 0$, thus such $\alpha$ and $\beta$ exist. This proves the result.
\end{proof}
\end{theorem}
With a similar argument to Theorem 10, we give the dual homogeneous distance of $C_{\mathcal{D}_2}$ and $C_{\mathcal{D}_3}$ in the following theorem, and we omit the proof here.
\begin{theorem}
 For $m\ge 2,$ the dual homogeneous distance $d_{hom}''$ and $d'''_{hom}$ of $C_{\mathcal{D}_2}$ and $C_{\mathcal{D}_3}$, respectively, is $2(p-1)p^{k-2}.$
\end{theorem}
\begin{remark}
In the case of $(p,k)=(2,2)$ with the defining set $\mathcal{D}_2$, we know from Theorem 11 that the dual homogeneous distance is 2, it is consistent with \cite[Theorem 7.2]{SL}.
\end{remark}

\section{Application of the linear codes to secret sharing schemes}
\subsection{The covering problem of linear codes}
\hspace*{0.6cm}The \emph{support} of a vector $c=(c_1,c_2,\ldots,c_n)\in\mathbb{F}_q^n$ is defined as $\{1\leq i\leq n|c_i\neq 0\}$. We say that a vector $x$ covers a vector $y$ if the support of $x$ contains the support of $y$. A \emph{minimal codeword} of a linear code $C$ is a nonzero codeword that does not cover any other nonzero codeword. The \emph{covering problem} of a linear code is to determine all the minimal codewords. However, in general determining the minimal codewords of a given linear code is a difficult task. In special cases, the Ashikhmin-Barg lemma \cite{AB} is very useful in determining the minimal codewords.
\begin{lemma}(Ashikhmin-Barg) In an $[n,k;q]$ code $C$, let  $w_{min}$ and $w_{max}$ be the minimum and maximum nonzero weights, respectively. If
\begin{equation}\label{1}
  \frac{w_{min}}{w_{max}}>\frac{q-1}{q},
\end{equation}
then all nonzero codewords of $C$ are minimal.
\end{lemma}
We can infer from there the support structure for the codes of this paper.
\begin{proposition}
 If one of the following two conditions satisfied
\begin{enumerate}
  \item[(1)] $m\geq 4$ even;
  \item[(2)] $m\geqslant 3$ odd, and $p\equiv 3 \pmod{4},$
\end{enumerate}
then all the nonzero codewords of $\Phi(C_{\mathcal{D}_1})$ are minimal.
\begin{proof} Following Theorem 1 and Lemma 7, we know $w_{min}=w_1$ and $w_{max}=w_3.$ Then we calculate $pw_1-(p-1)w_3$ as follows:
 \begin{eqnarray*}
  pw_1-(p-1)w_3&=& \frac{p-1}{p}p^{(k-1)(m+1)}\Big(\frac{p^m-1}{2}+\frac{p^{\frac{m}{2}}+1}{2}-p^{\frac{m}{2}+1}\Big).
 \end{eqnarray*}
Since $p$ is odd prime and $m\geq 4$ even, so $p^m+p^{\frac{m}{2}}-2p^{\frac{m}{2}+1}>0.$ The inequality (2) in Lemma 7 is satisfied.

 Likewise, take $w_{min}=w'_1$ and $w_{max}=w'_2.$ By a simple calculation we have $pw'_1-(p-1)w'_2>0$ for any odd prime $p\equiv 3 \pmod{4}$ and $m\geqslant 3$ odd.
\end{proof}
\end{proposition}

\begin{proposition}
 All the nonzero codewords of $\Phi(C_{\mathcal{D}_2})$ and $\Phi(C_{\mathcal{D}_3})$ introduced in Theorem $4$, for $m\geq 2$, are minimal.
\begin{proof}
Let $w_{min}=w''_1$ and $w_{max}=w''_2$ in the inequality (2) of Lemma 7, then we can check that the inequality holds for $m\geq 2$. The same discussion to $\Phi(C_{\mathcal{D}_3})$ in Theorem $4$ with $w_{min}=w'''_1$ and $w_{max}=w'''_2$.
\end{proof}
\end{proposition}

\subsection{Secret sharing schemes}

\hspace*{0.6cm} Secret sharing schemes (SSS) were first introduced by Blakley \cite{GR} and Shamir \cite{ASH}
at the end of the 1970s. Since then, many constructions have been proposed. Massey's scheme is a construction of
such a scheme which pointed out the relationship between the access structure and the minimal codewords of the dual code of the underlying code \cite{MM}.
See \cite{YJ} for a detailed explanation
of the mechanism of that scheme.
It would be interesting to know the dual Hamming distance (not the dual homogeneous distance),
as this would impact the SSS democratic or dictatorial character \cite{YC}.
We leave this as an open problem to the diligent reader.

\section{Conclusion}
\hspace*{0.6cm}This paper is devoted to the study of trace codes over a special finite chain ring of arbitrary depth. Using a character sum approach,
we have been able to determine their homogeneous weight distribution. Thus, several classes of $p$-ary two-weight codes,
and three-weight codes are obtained by the application of $\Phi,$ a linear Gray map defined in \cite{SS}. Furthermore, we have determined their dual
homogeneous distance. In particular, we have proved that the code $\Phi(C_{\mathcal{D}_1})$ and $\Phi(C_{\mathcal{D}_3})$ in the two-weight case, and the code
$\Phi(C_{\mathcal{D}_2})$ are optimal under some conditions. The codes we construct here have different parameters from those of the codes in \cite{RWM,SWY,SW}, thus
the obtained codes in the present paper are new, to the best of our knowledge. Moreover, when $(p,k)=(2,2)$ with the defining set $\mathcal{D}_2$ in this paper, the results coincide with \cite{SL}. Equivalently, this paper includes \cite{SL} as a special case.
Determining the dual Hamming distance of the considered codes is a challenging open problem, well-motivated by the secret sharing applications.

\end{document}